# Hydrodynamic Surrogate Models for Bio-Inspired Micro-Swimming Robots[*]

Ahmet Fatih Tabak, Serhat Yesilyurt[†]

*Mechatronics Program, Faculty of Engineering and Natural Sciences,*

*Sabanci University, Tuzla, 34956 Istanbul, Turkey*

**Summary**

Research on untethered micro-swimming robots is growing fast owing to their potential impact on minimally invasive medical procedures. Candidate propulsion mechanisms of robots are based on flagellar mechanisms of microorganisms such as rotating rigid helices and traveling plane-waves on flexible rods and parameterized by wavelength, amplitude, and frequency. For design and control of swimming robots, accurate real-time models are necessary to compute trajectories, velocities and hydrodynamic forces acting on robots. Resistive force theory (RFT) provides an excellent framework for the development of real-time six degrees-of-freedom surrogate models for design optimization and control. However, the accuracy of RFT-based models depends strongly on hydrodynamic interactions. Here, we introduce interaction coefficients that only multiply body resistance coefficients with no modification to local resistance coefficients on the tail. Interaction coefficients are obtained for a single specimen of *Vibrio Algino* reported in the literature, and used in the RFT model for comparisons of the forward-swimming component of the resultant velocities and body rotation rates against other specimens. Furthermore, CFD simulations are used to obtain forward and lateral velocities and body rotation rates of

---

bio-inspired swimmers with helical tails and traveling-plane waves for a range of amplitudes and wavelengths. Interaction coefficients are obtained from the CFD simulation for the helical tail with the specified amplitude and wavelength and used in the RFT model for comparisons of velocities and body rotation rates for other designs. Comparisons indicate that hydrodynamic models that employ interaction coefficients prove to be viable surrogates for computationally intensive three-dimensional time-dependent CFD models. Lastly, hydrodynamic models of bio-inspired swimmers are used to obtain optimal amplitudes and wavelengths of flagellar mechanisms, as a demonstration of the approach.

## 1. Introduction

Potential advantages of micro swimming robots can revolutionize the modern medicine: procedures such as kidney stone destruction, cleaning of clogged arteries, reaching tumors deep inside vital organs or retina restoration can be performed with minimal side-effects [1, 2]. Conventional mechanisms such as propellers cannot achieve propulsion at low Reynolds numbers in simple fluids, such as in micro fluids, as stated by Purcell's scallop theorem [3]. However, propulsion mechanisms of natural micro swimmers are viable candidates for propulsion of autonomous micro swimming robots [4]. Bio-inspired propulsion mechanisms have been demonstrated successfully in literature: a representative review is presented next.

Dreyfus et al. [5] demonstrated a novel artificial micro swimmer whose tail is chemically glued to a red blood cell and composed of magnetic particles attached to each other by DNA-protein strands. The tail was actuated by a combination of external dynamic magnetic fields to attain traveling planar waves and the swimmer moved in the opposite direction of the wave propagation. Yu et al. [6] carried out experiments with cm-scale flexible filaments, which mimic the whipping motion of spermatozoa, in viscous oils. The thrust force was measured by a strain gauge and the undulatory motion of the flexible tail was recorded by a camera; experimentally measured forces agreed well with the theoretical results of Wiggins and Goldstein [7]. Kosa et al. [8] proposed an actuation mechanism composed of piezoelectric laminates that deform and induce traveling plane waves. The propulsive force is calculated from the pendulum-like motion of the power cable that holds the robot.

Zhang et al. [9,10] manufactured artificial helical flagella as small as a few tens of microns long. Metal and polymer layers are deposited in the shape of a narrow tape and formed into a helix due to the tensile stress exerted on the inner layers during the manufacturing process. A magnetic bead of $4.5 \times 4.5 \times 0.2$ $\mu m^3$ is attached to one end of the artificial flagellum. In the presence of a rotating external magnetic field, the torque on the magnetic head enabled the rotation of the helical flagellum and the forward motion of the artificial swimmer [9,10]. Ghosh and Fischer [11] demonstrated the use of glancing angle deposition on the silicon wafer in an electron beam evaporator to obtain about a micron long helical screw-like structures with diameters of a few hundreds of nanometers. Helical structures are removed from the wafer, laid onto a surface and deposited by magnetic co-balt on one side. By means of a tri-axial Helmholtz coil, a rotating magnetic field is generated and modified by an open loop control scheme to navigate the microrobot on a preselected trajectory. Gao et al. [12] demonstrated swimming of a few microns long, flexible, Au/Ag/Ni nanowires, with an Au head and Ni tail linked by the partially dissolved silver bridge under an external rotating magnetic field. With a similar process presented in [12], Pak et al. [13] developed artificial swimmers with 1.5 µm-long Ni heads and 4 µm-long Ag tails and reported that the structure swims at about 20 μm/s under an external magnetic field rotating at 35 Hz. Tottori et al. [14] used 3-D direct laser writing process and physical vapor deposition to design and fabricate helical devices of varying lengths between 4and 65 μm. Authors demonstrated the corkscrew motion of helical structures in water and fetal bovine serum with rotating magnetic fields. Steering of the helical structure with a micro holder is achieved by changing the direction of the axis of rotation and utilized for transportation of microparticles by pick-and-place manipulation.

Hydrodynamic models of the propulsion of microorganisms date back to G.I. Taylor [15] who presented an analysis of the swimming of an infinite sheet, which deforms as traveling plane waves in an unbounded fluid; effects of the amplitude, wavelength, and frequency on the swimming speed of the sheet are formulated with the first-order perturbative approximation of the Stokes flow generated by the sheet. Gray and Hancock [16] presented the application of resistive force coefficients to calculate fluid forces due to the undulatory motion of slender filaments; force coefficients are obtained from the approximate solution of the fluid motion due to doublets and Stokeslets on the filament [17]. Sir James Lighthill [18] used the slender body theory for a swimmer with the helical tail based on the velocity field represented by Stokeslets and the corresponding distribution of point forces on the tail. Higdon [19] used a numerical integration method for the integrals approximated by Lighthill [18] to calculate the velocity of a swimmer with a spherical head and a helical tail, and reported the variation of the swimming velocity with the tail length, wavelength and amplitude given in dimensionless forms with respect to the diameter of the body. Lauga and Powers [20] present a comprehensive review of hydrodynamic models of swimming in micro scales.

There are a number of studies reported in the literature for artificial swimmers with flexible flagella, such as the structure developed by developed by Dreyfus et al. [5]. Here, we present a representative review: Roper et al. [21] modeled the artificial swimmer as a slender elastica driven by magnetic body torque the magnetostatic number, which is the ratio of magnetic and elastic forces. Gauger and Stark [22] presented a bead-spring model of the artificial swimmer to study the mean velocity and the efficiency of the swimmer as a function of the size of the particle, dimensionless Sperm number, which is the ratio of viscous forces to strength of the flexible filament, magnitude of the applied magnetic field and angular amplitude of the oscillating component of the field. According to results, the optimum size of the load is a compromise between the swimming velocity and the efficiency. Keaveny and Maxey [23] presented a particle-based numerical model of the artificial micro swimmer that consisted of paramagnetic beads as rigid spheres connected by inextensible flexible links, demonstrated that the model could be used to study corkscrew form of swimming driven by a rotating magnetic field, and obtained the velocity of the swimmer as a function of the Sperm number and the magnetostatic number . In [23], authors also presented a resistive force model for three-dimensional deformations of the flexible tail, and an analytical result for the swimming velocity at the low-frequency limit. Lastly, a slender-body model is presented for the swimmer developed by Pak et al. [13] by the authors, the model uses standard resistive force coefficients. In the model, bending stiffness of the tail is treated as a fitted parameter for a specified strength of the magnetic field and results agree well with experiments conducted for other magnetic field strengths.

Numerical solutions of the flow coupled with the equation of motion of the swimmer are carried out extensively in literature: the following is a representative review. Fauci and McDonald [24] presented a study of sperm motility near both rigid and elastic walls using the immersed boundary method to solve the two-dimensional time-dependent Navier-Stokes equations; authors report that the method proves to be useful especially for handling interactions with elastic walls. Ramia et al. [25] obtained instantaneous velocities of swimming of microorganisms from the solution of Stokes equations with the boundary element method in order to study hydrodynamic interactions between cells and solid boundaries as well as the interaction between the body and the tail of a cell. Goto et al. [26] employed the boundary element method for the solution of Stokes equations, calculated the velocity vector of a natural micro swimmer, compared their results with observations of actual swimmers, *Vibrio Alginolyticus*, and concluded that the BEM solutions agree reasonably well with observations. Qin et al. [27] studied the wall effects on a swimmer based on spermatozoa undergoing translations on a plane while fully submerged in a highly viscous fluid. Authors used immerse boundary method incorporating Navier-Stokes equations with Newton's second law to include swimmer's rigid-body accelerations, and computed the effective hydrodynamic interaction between the swimmer and parallel rigid plane walls based on the ratio of their half distance to wavelength.

Design and control of bio-inspired micro-swimming robots can benefit from accurate real-time hydrodynamic models that predict forward and lateral linear and angular velocities and trajectories of robots. It has been demonstrated that the resistive force theory (RFT) can be used to develop fast real-time models to predict the full three-dimensional trajectory of microswimmers [28,29,30]: in RFT-based models hydrodynamic forces on flagella are calculated from resistive force coefficients, and drag forces on bodies are obtained from analytical relationships for isolated objects, such as the well-known formula for a sphere. Models that use several variants of resistive force coefficients do not yield accurate predictions for even the forward velocity of a swimmer for all configurations of flagellar parameters such as the length of the tail, wavelength, and amplitude especially in the presence of a sizable body of the swimmer compared to the tail [19,28,30,31]. The hydrodynamic interaction between the body and the tail is one of the key phenomena which are not included properly in RFT models.

Hydrodynamic interactions are studied for organisms near planar walls, e.g., [6,29], and for two or more organisms swimming together, e.g., [32,33]; however, the influence of the flagellar motion on the body of swimmers has not been addressed thoroughly in literature to the best of our knowledge. Lighthill included the effect of the cell body on the slender-body-theory-based calculations of the velocity of the swimmer and concluded that the correction is very small compared to an isolated infinite flagellum [18]; Chattopadhyay and Wu demonstrated that Lighthill's correction is very small for micro swimming species such as *Vibrio Algino* [30]. Furthermore, the hydrodynamic interaction between the body and the flagellum was studied numerically by Ramia et al. [25]. Authors concluded that the presence of the cell body does not alter the flagellar propulsion force as significantly as the flagellar force alters the total drag force on the cell. However, authors did not provide detailed results for the effect of flagellar parameters on the drag force on the body attached to an actuated flagellum, such as a helix, compared to an isolated body with the same shape and size.

Hydrodynamic forces on tails are obtained from the integration of local forces in tangent and perpendicular directions to the motion and expressed by resistive force coefficients over the tail; force coefficients can be calculated from analytical formulas available in the literature, such as from Lighthill's slender body theory. Body resistance coefficients are known for isolated objects such as spheroids in unbounded fluid media: for example the resistance coefficient is $F_i / U_i = -6\pi\mu a$ for the hydrodynamic force $F_i$ acting on an isolated spherical object of radius $a$ moving with velocity $U_i$ in an unbounded fluid of viscosity $\mu$ in the $i^{th}$ direction; and the resistance is the same for all directions. In the presence of an actuated tail attached to the spherical body, it is clear that the symmetrical drag relationship breaks. We propose hydrodynamic interaction coefficients, $\gamma_i$, which scale the resistance coefficient for the motion in the $i^{th}$ direction, namely $F_i / U_i = -\gamma_i (6\pi\mu a)$. Hydrodynamic interaction coefficients are different for each direction due to the rotation of the helical tail, which breaks the symmetry of the flow over the spherical body. Calculation of hydrodynamic interactions is extremely difficult analytically; however, an experiment or a CFD simulation can be performed for a fixed representative design once to obtain hydrodynamic interaction coefficients.

The inverse problem is solved once with a single CFD-simulation for a representative design with fixed values of the design variables such as amplitude and wavelength of the helical waves to obtain unknown coefficients. The method is somewhat ad-hoc in the prediction of hydrodynamic interaction coefficients for the sake of improving the accuracy of the RFT model which takes seconds to run instead of hours in the case of three-dimensional CFD simulations subject to free-swimming constraints. The RFT-based model, which is presented here, serves as a surrogate for accurate numerical models, such as CFD simulations, and can be used as a real-time model in model-based control and design optimization studies to search alternative designs in the neighborhood of the representative design. Therefore the main question is the extent of the neighborhood in which the RFT-based model retains its accuracy. Lastly, the approach is valid only for design optimization problems that can be represented by a finite number of design variables, which are used to parameterize a given waveform of the tail and the geometry of the body, and is not an alternative for generalized shape optimization strategies, for example as recently presented by Keaveny et al. [34].

The RFT-based hydrodynamic model is validated with measurements of Goto et al. [26] for a group of species of microorganisms with the varying body and tail dimensions, and with three-dimensional time-dependent CFD simulation experiments for swimmers with designs other than the one used to obtain interaction coefficients. Furthermore, the validated hydrodynamic model is used to obtain optimal efficient tail parameters for desired operations such as efficiency and speed as a demonstration.

## 2. Methodology

### 2.1 Hydrodynamic model

The time-dependent trajectory of a two-link, micro-swimmer is obtained from the equation of motion, which balances forces on the swimmer's body and the tail:

$$F_b + F_t = 0, \qquad (1)$$

where $F = [\mathbf{F}', \mathbf{T}']'$ is the generalized force vector, $\mathbf{F}$ and $\mathbf{T}$ are the force and torque vectors, "'" is the transpose, and subscripts $b$ and $t$ refer to the body and the tail. For simplicity, we assume that the body of the swimmer is a blunt object such as a sphere, and the flexible tail is subject to a motion that generates propulsion force in viscous flows, such as the rotation of a helix, or traveling-plane waves on a slender rod as commonly observed among micro swimming organisms.

For creeping flows, at low Reynolds numbers, equations of motion can be cast in a linear system of equations relating the generalized force and velocity vectors by means of the resistance matrix, $\mathbf{B}_i$, as follows:

$$F_i = -\mathbf{B}_i \mathbf{V}_i. \qquad (2)$$

Here, $i = \{b, t\}$, $\mathbf{B}$ is the resistance matrix, $\mathbf{V} = [\mathbf{U}', \mathbf{\Omega}']'$ is the generalized velocity vector, $\mathbf{U}$ and $\mathbf{\Omega}$ are translational and rotational velocity vectors respectively.

The resistance matrix for the rigid body of the swimmer, $\mathbf{B}_b$, is simpler than the resistance matrix of the tail and obtained from the linear and rotational resistance of the body and can be considered as a combination of four subcomponents which relate linear and angular velocities to forces and torques:

$$\mathbf{B}_b = \begin{bmatrix} \mathbf{D}_N & \mathbf{E} \\ \mathbf{E}' & \mathbf{D}_R \end{bmatrix},\qquad(3)$$

where matrices $\mathbf{D}_N$ and $\mathbf{D}_R$ are 3×3 diagonal matrices that correspond to the translational and rotational resistance of the body, $\mathbf{E}$, which contains nonzero elements if body center of mass and swimmer center of mass are far apart along any direction.

For a spherical isolated body in an unbounded fluid, each diagonal element of $\mathbf{D}_N$ is $6\pi\mu r_b$ and each diagonal element of $\mathbf{D}_R$ is $8\pi\mu r_b^3$, where $\mu$ is the dynamic viscosity and $r_b$ is the radius of the spherical body. Drag coefficients for generalized ellipsoids and other body shapes are also known [35,36,37]. However, even for a simple body such as a sphere, resistance matrices $\mathbf{D}_N$ and $\mathbf{D}_R$ must be modified due to the motion of the tail attached to the body as well as for flows inside channels and nearby boundaries [38,39]. For instance, the respective interaction coefficients can be introduced into the resistance matrix of a spherical body as follows:

$$\mathbf{D}_N = 6\pi\mu r_b \begin{bmatrix} \gamma_{N,s} & 0 & 0 \\ 0 & \gamma_{N,q} & 0 \\ 0 & 0 & \gamma_{N,r} \end{bmatrix},\qquad(4)$$

where $\gamma_{N,i}$ is the coefficient modifying the spherical drag in the $i^{\text{th}}$ direction.

In effect, the interaction coefficients are only applied to the body resistances: due to the linearity of the equation of motion, Eq. (1), the relative effect of the hydrodynamic inter-actions can be applied to either the body of the swimmer or its tail. Interaction coefficients need to be estimated well in order to ensure the accuracy of the hydrodynamic model.

Time-dependent resistance matrix of the tail, $\mathbf{B}_t$ in (2), is obtained from integration of local forces:

$$\mathbf{B}_t = \int_0^L \begin{bmatrix} \mathbf{RCR'} & -\mathbf{RCR'S} \\ \mathbf{SRCR'} & -\mathbf{SRCR'S} \end{bmatrix} ds, \qquad (5)$$

where $L$ is the apparent length of the tail in the **s**-direction, **S** is the skew-symmetric matrix that corresponds to the cross product with the position vector on the tail, **R** is the rotation matrix between the local Frenet-Serret coordinates, **t-b-n**, and the **s-q-r** coordinates of the swimmer (Fig. **1**), and formed by local tangential, **t**, bi-normal, **b**, and normal, **n**, vectors [40]:

$$\mathbf{R} = \begin{bmatrix} \mathbf{t}(s,t) & \mathbf{b}(s,t) & \mathbf{n}(s,t) \end{bmatrix}. \qquad (6)$$

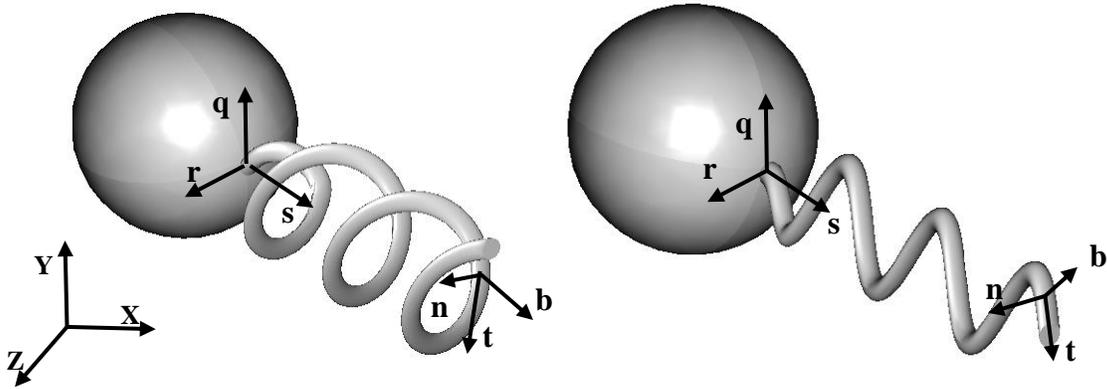

**Fig. 1.** Swimmer with a rotating helical tail (left), and with traveling plane waves (right) and corresponding frames of reference: **XYZ** is the stationary frame; **sqr** is located at the joint and translates with the body; and **tbn** is the local time-dependent Frenet-Srenet coordinates of an arbitrary location on the tail.

The local resistance matrix at a given position on the tail, **C** in Eq. (5), is a diagonal matrix that consists of the local resistance coefficients in the tangent, $c_t$, bi-normal and normal directions, $c_n$. Local resistance coefficients are the same in the bi-normal and normal directions as both are perpendicular to the tangential direction.

Accurate calculation of the resistance coefficients is extremely difficult. Lighthill derived resistance coefficients from the distribution of Stokeslets and point forces on an infinite helix in an unbounded fluid [18]. A number of simplifying assumptions are used in the derivation of resistance coefficients. The local normal and tangential components of the resistive force coefficients are obtained from as follows [18]:

$$c_n = \frac{4\pi\mu}{-\ln\varepsilon + (2\alpha^2 - 1)A_1 + 2(1-\alpha^2)A_2}, \tag{7}$$

and

$$c_t = \frac{2\pi\mu}{-\ln\varepsilon - 0.5 + \alpha^2 A_1 + (1-\alpha^2)A_2}. \tag{8}$$

Here α is the ratio between apparent and actual lengths of the tail; ε is given by a relationship based on tail's radius, $a$, $\alpha$, and wavelength, λ: $\varepsilon = 5.2 a\alpha/\lambda$. $A_1$ and $A_2$ are periodic integrals of functions of assumed local flow fields and specified in [18] as follows (also please refer to the **Appendix** section):

$$A_{\{1,2\}} = \ln\varepsilon + \int_\varepsilon^\infty \frac{\{\theta\sin\theta, \sin^2\theta\}}{\left[\alpha^2\theta^2 + 2(1-\alpha^2)(1-\cos\theta)\right]^{3/2}} \, d\theta. \tag{9}$$

Local velocity on the tail is the summation of the swimmer's net velocity and the motion of the tail with respect to body coordinates, i.e.,

$$\mathbf{U}_t = \mathbf{U}_b + \hat{\mathbf{u}}, \tag{10}$$

where $\hat{\mathbf{u}}$ is the local velocity on the tail, and can be obtained from the deformation or the rotation of the tail:

$$\hat{\mathbf{u}} = \frac{d\mathbf{P}}{dt}. \tag{11}$$

For instance, for a left-handed helical tail as shown in Fig. **1**, the position vector, $\mathbf{P} = [s\ q\ r]'$, is specified in the body coordinate frame as follows:

$$\mathbf{P} = \begin{bmatrix} s \\ q(s,t) \\ r(s,t) \end{bmatrix} = \begin{bmatrix} s \\ b_s(s)\cos(ks - \omega t) \\ -b_s(s)\sin(ks - \omega t) \end{bmatrix}, \tag{12}$$

where $k$ is the wave number and $b(s)$ is the local radius of the helix, which is modified with a ramp function to ensure a fixed connection with the body (Fig. **1**), e.g., $b_s(s) = 10b(s/L)$ for $s/L < 0.1$ and $b_s(s) = b$ for $(s/L) \geq 0.1$.

In the case of plane-wave deformation, the local position is specified by the $q$-displacement and the $r$-displacement is set to zero. For an arbitrary actuation mechanism, the velocity of the tail can be calculated from Eq. (10) once the local position vector on the tail is specified.

Forces on the tail can be decomposed into propulsion and drag forces; according to Eqs. (2) and (10), then, we have:

$$\begin{bmatrix} \mathbf{F}_t \\ \mathbf{T}_t \end{bmatrix} = -\mathbf{B}_t \begin{bmatrix} \mathbf{U}_t \\ \mathbf{\Omega}_t \end{bmatrix} = -\mathbf{B}_t \left( \begin{bmatrix} \mathbf{U}_b \\ \mathbf{\Omega}_b \end{bmatrix} + \hat{\mathbf{u}} \right) = -\mathbf{B}_t \begin{bmatrix} \mathbf{U}_b \\ \mathbf{\Omega}_b \end{bmatrix} - \begin{bmatrix} \mathbf{F}_p \\ \mathbf{T}_p \end{bmatrix}. \tag{13}$$

The first term in the right-hand-side of the last equation in Eq. (13) is the total drag force on the tail due to its motion with the body, and the second term is the propulsion force and torque generated by the tail due to its motion relative to the body, i.e., $\mathbf{B}_t \hat{\mathbf{u}}$, and obtained from Eq. (5) as follows:

$$\begin{bmatrix} \mathbf{F}_p \\ \mathbf{T}_p \end{bmatrix} = \int_0^L \begin{bmatrix} \mathbf{RCR}\hat{\mathbf{u}} \\ (\mathbf{P} - \mathbf{P}_{com}) \times (\mathbf{RCR}\hat{\mathbf{u}}) \end{bmatrix} ds, \tag{14}$$

where $\mathbf{R}$ is given by Eq. (6), $\mathbf{C}$ is the local resistance matrix. $\mathbf{P}_{com}$ is the position of the center of mass in **sqr** coordinates.

Substituting (2), (5) and (13) into (1), one obtains the instantaneous velocity vector of the swimmer in the **sqr** coordinates:

$$\begin{bmatrix} \mathbf{U}_b \\ \mathbf{\Omega}_b \end{bmatrix} = (\mathbf{B}_b + \mathbf{B}_t)^{-1} \begin{bmatrix} \mathbf{F}_p \\ \mathbf{T}_p \end{bmatrix}. \tag{15}$$

In the case of traveling plane-waves, the propulsion force and torque are only due to the **q**-component of $\hat{\mathbf{u}}$, **r**-component of $\mathbf{U}_b$, whereas **s**- and **q**-components of $\mathbf{\Omega}_b$ are zero.

In order to obtain the velocity of the swimmer in the lab frame for control studies, the rotation matrix, $\mathbf{R}_L$, between the **sqr** and **XYZ** frames (see Fig. **1**) must be calculated either explicitly from Euler angles or from quaternion transformations [41]. In order to alleviate the representation problem, we implement the latter and obtain the velocity vector in the lab frame:

$$\begin{bmatrix} \mathbf{U}_b^{\mathbf{XYZ}} \\ \mathbf{\Omega}_b \end{bmatrix} = \begin{bmatrix} \mathbf{R}_L & \mathbf{0} \\ \mathbf{0} & \mathbf{I} \end{bmatrix} \begin{bmatrix} \mathbf{U}_b \\ \mathbf{\Omega}_b \end{bmatrix}. \tag{16}$$

Once the velocity vector in the lab frame is obtained, the position of the swimmer is obtained kinematically, for example, with a Runge-Kutta scheme. The quaternion for the rotation matrix $\mathbf{R}_L$ is also part of the integration scheme to keep track of the orientation of the swimmer. Typical simulation time for a swimmer with a helical tail is less than a second for the full rotation of the tail on a high-end mobile workstation.

**2.2 CFD model**

Computational fluid dynamics (CFD), which provides a numerical solution of three-dimensional time-dependent Navier-Stokes equations, is used to compute reliably fluid forces especially at low Reynolds number flows. In order to model the motion of a swimming robot in an unbounded fluid, here, we use a relatively large channel around the swimmer with the diameter as large as ten times the diameter of the body, and length five times the total length of the swimmer with negligible distortion to the flow field nearby the swimmer.

Fluid forces are calculated from the finite-element method solution of incompressible Navier-Stokes equations in the moving domain due to the motion of the tail and the overall swimmer. Arbitrary-Lagrangian-Eulerian (ALE) scheme [42] is used in order to handle the deforming mesh. Equations are nondimensionalized with the diameter of the body, $D_b$, as the length scale and $2\pi/\omega$ as the time-scale; hence the velocity scale is $\omega D_b/2\pi$, which varies linearly with the frequency, and the scaling Reynolds number is $\mathrm{Re} = \rho\omega D_b^2/2\pi\mu$. A complete list of variables used in the representative, base-case design is shown in Table 1.

Hydrodynamic forces on the swimmer are computed from the integration of the stress distribution over the surface of the swimmer and set to zero as a set of constraint equations in order to obtain forward and lateral velocities and body rotation rates from no-slip moving boundary conditions on the swimmer:

$$\begin{bmatrix} \mathbf{F}_t \\ \mathbf{T}_t \end{bmatrix} + \begin{bmatrix} \mathbf{F}_b \\ \mathbf{T}_b \end{bmatrix} = \begin{bmatrix} \int\limits_{Tail+Body} \tau \mathbf{n} \, dA \\ \int\limits_{Tail+Body} (\mathbf{x} - \mathbf{x}_{com} \times \tau) \mathbf{n} \, dA \end{bmatrix} = 0 . \tag{17}$$

Here, $\tau$ is the fluid stress tensor, **n** is the time-dependent three-dimensional local surface normal, **x** is the position vector, and $\mathbf{x}_{com}$ is the position of the center of mass, which is assumed to be the geometric center of the spherical body.

Full translations of the rigid body and the rotation of the body along the **s**-axis are obtained for swimmers with helical tails from (17). Specifically, the first row of (17) is the constraint equation for the forward velocity of the swimmer, the second row for the **q**-velocity, third row for the **r**-velocity, and the fourth row is for the angular velocity of the body around the **s**-axis. CFD simulations are carried out for two types of flagellar propulsion mechanisms: the first one is the left-handed helical tail rotating in the positive direction with respect to the **s**-coordinate as shown in Fig. **1**a; and the second one is for the tail with traveling waves in the **q**-**s** plane as shown in Fig. **1**b. Independent rotation of helical tails is observed in microorganisms and demonstrated as an effective mechanism with large-scale experiments in viscous fluids [43,44]. Traveling waves are used to simplify the deformation of flexible filaments; actual deformations can be modeled with elastic properties to replace the amplitude and wavelength, which are used as independent variables here. Spherical body is chosen for its simplicity and well-known drag coefficients. The approach, which is presented here, can easily be extended to study magnetized artificial structures with arbitrary body shapes and flexible filaments without loss of generality.

Commercial software, COMSOL Multiphysics [45], which is based on the finite-element method, is used to perform the simulations with the second order Lagrangian tetrahedral elements. For all simulations, 300,000 degrees-of-freedom is used. Linear system of equations is solved with the PARDISO linear solver and a second order backward difference formula with variable time-stepping for the numerical integration in time (maximum time step is set to 0.0025). Simulations require up to twenty hours on a high-end workstation in order to complete two full periods of the wave propagation (helical or planar) on the tail depending on its geometric parameters.

**Table 1**
Base case parameters and their dimensionless values for swimmers in CFD simulations.

| Parameter Name | Dimensionless Value |
| --- | --- |
| Radius of the spherical body, $r_b$ | 0.5 |
| Chord radius of the tail, $r_t$ | 0.05 |
| Apparent length of the tail, $L$ | 2 |
| Apparent wavelength, $\lambda$ | 2/3 |
| Wave amplitude, $b$ | 0.1 |
| Actuation frequency of the tail, $\omega/2\pi$ | 1 |
| Fluid density, $\rho$ | 1 |
| Scaling Reynolds number, $Re = \rho\omega D_b^2/2\pi\mu$ | $10^{-2}$ |
| Cylindrical channel length, $L_{ch}$ | 10 |
| Cylindrical channel diameter, $2r_{ch}$ | 10 |
| Spherical body resistances, $\{6\pi\mu r_b, 8\pi\mu r_b^3\}$ | {942.5, 314.2} |

## 3. Results

### 3.1 Validation of the hydrodynamic model with measurements

Hydrodynamic surrogate model is compared against measurements reported in the literature. Resistive force coefficients from Lighthill's slender body theory [18], which are given by (7) and (8), are used for the helical tail in the hydrodynamic model in this part.

Goto et al. [26] measured forward velocity and body rotation rates for a number of specimens of *Vibrio Alginolyticus*, whose dimensions and tail rotation rates vary individually. Authors could not measure the frequency of rotations of the helical tail, due to the relatively high frequency of tail rotations compared to body rotations, and used a boundary-element method (BEM) model to calculate the tail-rotation frequency from the measured frequency of body rotations. Table 2 shows reported geometric parameters of individual organisms; for all cases, the radius of the tail is 16 nm, the wavelength of the helical waves is 1.37 μm and the amplitude (helical radius) is 0.1487 μm [26]. In effect, amplitude and wavelength are fixed in these measurements, and tail lengths and body dimensions vary.

Translational and rotational resistance coefficients of the body in the swimming direction (**s**-direction in Fig. **1**) are calculated with the use of drag coefficients for oblique spheroids from [36]:

$$\mathbf{D}_{N,s} = \Gamma_{N,s} 4\pi\mu r_s \big/ \big(\log(2r_s / r_q) - 0.5\big), \tag{18}$$

and

$$\mathbf{D}_{R,s} = \Gamma_{R,s} (16/3) \pi\mu r_s r_q^2. \tag{19}$$

In (18) and (19), $r_{\{s,q\}}$ are the radii of the body in the **s** and **q**-directions respectively, and $\Gamma_{\{N,R\},s}$ are hydrodynamic interaction coefficients, that correspond to variations in body drags from the ideal case for isolated spheroids in an unbounded fluid. In essence, interaction coefficients quantify variations in translational and rotational body drag coefficients due to the flow field realized by the rotating tail attached to the body. If interaction coefficients in (18) and (19) are set to unity, translational and rotational drag factors, $\mathbf{D}_{N,s}$ and $\mathbf{D}_{R,s}$, would be those of isolated spheroids in infinite media.

Time-averaged forward velocity and the body-rotation rate of natural swimmers are calculated from (15) and compared with the measurements of Goto et al. [26] in Fig. **2**. There is a significant discrepancy between the measurements and model results when interaction coefficients are set to unity, i.e., for $\Gamma_{\{N,R\},s} = 1$: maximum error is found to be 87% in the average forward velocity for specimen G, and 47.2% in the body-rotation rate for specimen B.

**Table 2**
Geometric parameters of V. *Alginolyticus* specimens.

| Specimen | Frequency (Hz) | Tail Length (μm) | Body **s**-semi-axis, $r_s$ (μm) | Body **q** and **r**-semi-axes $r_q$ (μm) |
|---|---|---|---|---|
| A | 187.70 | 4.89 | 1.885 | 0.415 |
| B | 123.20 | 4.90 | 1.320 | 0.380 |
| C | 73.95 | 5.24 | 1.380 | 0.405 |
| D | 244.70 | 5.19 | 1.975 | 0.400 |
| E | 126.20 | 5.03 | 1.785 | 0.405 |
| F | 220.10 | 5.07 | 2.260 | 0.380 |
| G | 477.10 | 4.87 | 2.280 | 0.410 |

Values of two interaction coefficients, $\Gamma_{N,s}$ and $\Gamma_{R,s}$, can be determined from the solution of the inverse problem for observed values of the forward velocity and the body rotation rate of a selected swimmer as the representative design. Here, specimen C is used as the representative design, and the interaction coefficients in translational and rotational drag relationships given by Eqs. (18) and (19) are calculated as $\Gamma_{N,s} = 2.37$ and $\Gamma_{R,s} = 1.49$ respectively from the solution of the inverse problem. As shown in Fig. **2**, the agreement between the hydrodynamic model and measurements is very good with updated resistance coefficients of the body including interaction coefficients: maximum error is 8.2% in the average forward velocity for specimen G, and 6.5% in the body rotation rate for specimen F. Despite that specimens have different body dimensions, tail lengths and tail rotation frequencies (Table 2), interaction coefficients obtained from the solution of the inverse problem for an arbitrarily selected specimen work very well other specimens as well. Thus, once the resistance coefficients of the body are obtained accurately, the hydrodynamic model would perform sufficiently well in subsequent analyses.

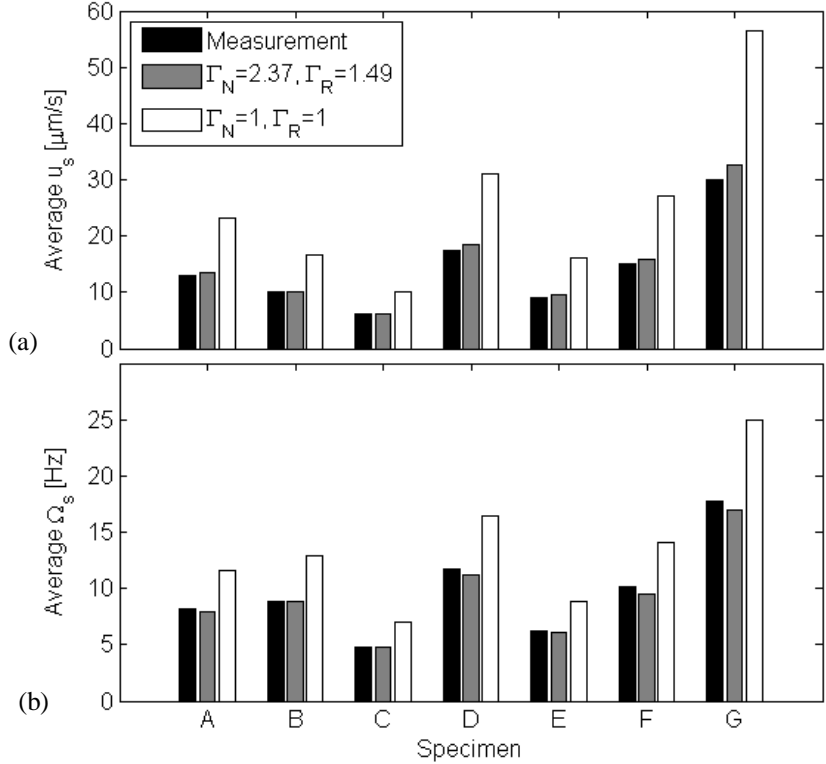

**Fig. 2.** Comparisons of the time-averaged forward velocity (a) and angular velocity of the body (b); between the measurements reported by Goto et al. [26], the model for unmodified ($\Gamma_N = \Gamma_R = 1$), and corrected body drags ($\Gamma_N = 2.37$ and $\Gamma_R = 1.49$) as in Eq. (18) and Eq. (19).

It is reasonable to expect that hydrodynamic interactions between the body and the tail would have an effect on the resistance force coefficients of the tail as well. In effect, the linearity of the equation of motion, which consists the force-free swimming condition given by Eq. (1) and the resistance relationship between the forces and velocities given by Eq. (2), allows that hydrodynamic interactions can be included in the resistance matrix of only one component, either the body or the tail. Furthermore, results of previous numerical studies show that the total drag force on the tail is not affected by the choice of body as much as the total drag force on the whole swimmer is affected [25].

## 3.2 CFD simulations

### *3.2.1 Estimation of hydrodynamic model parameters*

<u>Two sets of resistive force coefficients</u> are used for tails in the hydrodynamic model: the first set is by Lighthill [18] and obtained from Eqs. (7) and (8); and the second set is obtained from the CFD simulation for a stationary swimmer with a rotating helical tail. The helical radius (amplitude) of the tail is set to 0.1 and the wavelength to 2/3 as the base case design. A complete list of base-case design parameters and their values are provided in Table 1.

Stationary swimmer in the CFD simulation is not subject to the force-free swimming constraints given by Eq. (17), thus the rotating left-handed helical tail of the swimmer generates a net propulsion force in the opposite direction of the rotation; the propulsion force and the torque on the swimmer's body can be calculated from the integration of the fluid stress field over the tail in the CFD model. Then, integrations on the right-hand-side of Eq. (14) are carried out explicitly only in the swimming direction (**s**-direction in Fig. **1**) to obtain a closed-form expression for the force and the torque generated by the tail and the rotation and the translation velocities as follows:

$$\begin{bmatrix} \bar{F}_{t,s} \\ \bar{T}_{t,s} \end{bmatrix} = -\alpha L \left\{ \begin{bmatrix} b^2 k^2 & 1 \\ b^2 k & -b^2 k \end{bmatrix} \bar{u}_s + \begin{bmatrix} k & -k \\ 1 & b^2 k^2 \end{bmatrix} b^2 \omega \right\} \begin{bmatrix} c_n \\ c_t \end{bmatrix}, \qquad (20)$$

where $\alpha = (1+b^2 k^2)^{-1/2}$ is the ratio of the apparent length of the helix to the actual rod length of the tail, $b$ is the helical radius, which is 0.1 for the base case, $k$ is the wavenumber, which is 3 for the base case, $\bar{u}_s$ is the average swimming speed, which is zero for the stationary swimmer, and $\omega$ is the frequency of rotations of the tail. Once the left-hand-side of Eq. (20) is computed from the CFD model for the stationary swimmer, resistive force coefficients, $c_n$ and $c_t$ are, then, easily calculated as 995.5 and 775.2, respectively. Arguably, the constant pair of force coefficients, which are obtained from the CFD simulation, incorporates realistic flow conditions such as the finite length and radius of the tail and the trailing-edge force due to the motion of the tip of the tail, which are not taken into account in the derivation of the resistance coefficients from the slender body theory [18].

According to Eqs. (7) and (8), resistive force coefficients vary with the parameter α, which is the ratio of the chord length of the tail to its apparent length and varies with the amplitude and wavelength. Fig. **3**a shows the variation of the $c_n/c_t$ ratio with respect to the total number of waves and Fig. **3**b shows the variation of the ratio with respect to amplitude for helical tails and traveling plane waves. The constant $c_n/c_t$ ratio for the pair, which is obtained from the CFD simulation for the base case, is also shown on the plots in Fig. **3**. For traveling-plane-wave tails, we used the wavelength-averaged value of α, the ratio of the chord length to apparent length, as it varies locally unlike the ratio for the helical tail, which remains constant independent of the local position on the tail.

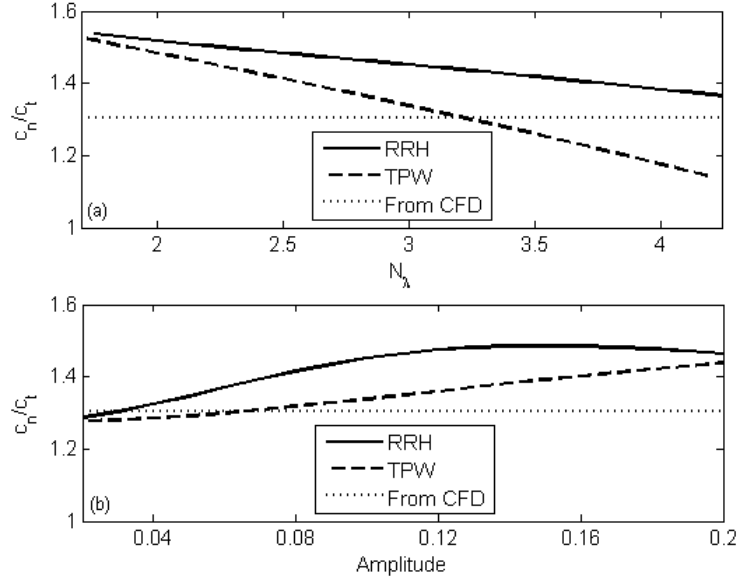

**Fig. 3.** The ratio of the resistive force coefficients obtained from Lighthill's slender-body-theory [18], as a function of the number of waves (a), and amplitude (b) for rotating rigid helical (RRH) tails and traveling plane waves (TPW).

Resistance coefficients for the body are obtained from the well-known drag coefficients of spherical objects multiplied by translational and rotational hydrodynamic interaction coefficients in the $k^{\text{th}}$ direction, $\gamma_{N,k}$ and $\gamma_{R,k}$ respectively, and used as diagonal factors in the body resistance sub-matrices in (3) as follows:

$$\mathbf{D}_N = 6\pi\mu r_b \begin{bmatrix} \gamma_{N,s} & 0 & 0 \\ 0 & \gamma_{N,q} & 0 \\ 0 & 0 & \gamma_{N,r} \end{bmatrix}, \qquad (21)$$

and

$$\mathbf{D}_R = 8\pi\mu r_b^3 \begin{bmatrix} \gamma_{R,s} & 0 & 0 \\ 0 & \gamma_{R,q} & 0 \\ 0 & 0 & \gamma_{R,r} \end{bmatrix}. \qquad (22)$$

Interaction coefficients account for the hydrodynamic effect of the tail's motion on the body's resistance coefficients, which are the diagonal elements of the resistance matrix of given in (3). Off-diagonal elements of the body resistance matrices can be used to account for more general interactions between the directions of body's motion, for example the well-known Magnus effect, which is recently observed for microparticles at very low Reynolds numbers [46], can be described as the force in the **q**-direction due to the motion of the body in the **s**-direction and the rotation in the **r**-direction (see Fig. **1** for directions). Moreover, a strictly-diagonal form of the body resistance matrix, which is considered here, can be viewed as the result of the diagonalization of a general form that includes all hydrodynamic interactions and is currently under investigation [47].

For helical propulsion, it is assumed that in addition to the interaction coefficient in the swimming direction, $\gamma_{N,s}^{Helix}$, only a single translational resistance for the body in lateral directions, i.e., $\gamma_{N,q}^{Helix} = \gamma_{N,r}^{Helix}$, is necessary; and only one coefficient is necessary for the rotation in the swimming direction, $\gamma_{R,s}^{Helix}$, as body rotations in other directions are not calculated in the model for helical tails for simplicity.

Interaction coefficients for the spherical body of the free swimmer that corresponds to the base-case representative design are calculated directly from the ratio of forces and velocities obtained from the CFD simulation. The time-dependent forward velocity of the swimmer is nearly constant varying within 0.6% of its average value, −0.038. The net hydrodynamic drag force in the swimming direction of the spherical body of the swimmer is obtained as 81.7, which corresponds to 2.28 times the well-known drag force on spher-ical objects. Therefore the interaction coefficient in the swimming direction is obtained as, $\gamma_{N,s}^{Helix,CFD} = 2.28$. Similarly, the angular velocity of the swimmer is almost constant varying within 0.2% of its time-averaged value, which is obtained as −0.4; the torque exerted on the spherical body is 1.09 times its well-known value for spherical objects and sets the value interaction coefficient for rotations in the swimming direction as $\gamma_{R,s}^{Helix,CFD} = 1.09$.

Lateral (**q**- and **r**-directions in Fig. **1**) velocities and forces are both sinusoidal in time with zero mean and amplitude of 0.015 and 7.235 respectively. The phase between the waveforms of lateral velocities and forces is equal to $\pi/2$. The ratio of the amplitudes of the lateral forces and the lateral velocities is 0.51 times the spherical drag; however, for lateral directions, we use the interaction coefficient from the solution of the inverse problem and obtained as $\gamma_{N,q}^{Helix,CFD} = \gamma_{N,r}^{Helix,CFD} = 1.24$.

Actual values of interaction coefficients vary with the choice of resistive force coefficients used for the tail since the effect of hydrodynamic interactions between the body and the tail is evaluated by the interaction coefficients applied only to body resistance. Therefore, for the resistive force coefficients obtained from Liqhthill's slender-body-theory [18], a new set of interaction coefficients are necessary. In this case, we solve the inverse problem for, i.e., Eq. (15), for already calculated velocities to obtain interaction coefficients for the body resistance matrices, which are calculated as $\gamma_{N,s}^{Helix,SBT} = 3.35$, $\gamma_{N,q}^{Helix,SBT} = \gamma_{N,r}^{Helix,SBT} = 1.1$, and $\gamma_{R,s}^{Helix,SBT} = 0.85$.

<u>Flagellar propulsion with traveling plane waves (TPW)</u>, in essence, can be considered as a special case of the helical propulsion since the deformation of the tail in the **r**-direction is set to zero in Eq. (12). Therefore, it is assumed that resistive force coefficients obtained from Eq. (20) for the stationary swimmer with the helical tail should perform reasonably well here. In this case, interaction coefficients for body resistances are required for the forward motion of the swimmer in the **s**-direction, $\gamma_{N,s}^{TPW}$, lateral motion of the swimmer in the **q**-direction, $\gamma_{N,q}^{TPW}$, and the rotation of the body in the **r**-direction, $\gamma_{R,r}^{TPW}$ (see Fig. **1**b for the directions). The interaction coefficient in the swimming direction is calculated from the ratio of the time-averaged force and the time-averaged velocity in that direction, obtained from the CFD simulation of the free swimmer with the traveling-plane-wave tail whose amplitude and wavelength are set to the base-case values, 0.1 and 2/3 respectively; the calculated value of the interaction coefficient is obtained as, $\gamma_{N,s}^{TPW,CFD}=2.21$. This value is very close to the one obtained for the swimmer with the helical tail.

Similarly, from the ratio of the amplitudes of the lateral force and the lateral velocity, which are zero in average, the lateral interaction coefficient is obtained as $\gamma_{N,q}^{TPW,CFD}=3.14$. Lastly, the interaction coefficient for the rotational resistance of the body in the **r**-direction perpendicular to the plane of propagating wave is obtained from the amplitude ratio of the torque and the angular velocity in that direction as $\gamma_{R,r}^{TPW,CFD}=0.45$.

Interaction coefficients are also calculated from the solution of the inverse problem for both sets of resistive force coefficients, from the CFD simulation for the stationary swimmer with a helical tail and from Eqs. (7) and (8); results are presented in Table 3. It is somewhat surprising to see that the interaction coefficient for the **r**-rotation of the swimmer is negative. We suspect that the result is an artifact of using only the diagonal com-ponents of the body resistance matrix. In effect, the **r**-rotation of the swimmer is strongly linked with the **q**-translation of the swimmer due to the strong coupling between the **r**-torque on the swimmer and the **q**-force on the tail (see Fig. **1**b for directions) and the fact that the center of mass of the swimmer coincides with the center of the spherical body. In essence, having a very large interaction coefficient in the **q**- translation and a negative one for the **r**-rotation could be the result of more complex interactions between the two modes of the motion. For example, the Magnus effect, although does not apply here but observed in microflow conditions recently [46], can be represented by a negative lateral resistance due to the rotation and forward motion of a spherical body. Our efforts continue to investigate this matter further to elucidate the extent of hydrodynamic interactions between the body and the tail.

**Table 3**
Interaction coefficients for body the resistance matrices of the spherical body given by Eqs. (21), (22), and (23).

| Propulsion type | Tail resistance coefficient | Body drag factor |
|---|---|---|
| Helical | From (7) and (8) | $\gamma_{\{N,s;N,q;R,s\}}^{Helix,SBT} = \{3.35; 1.1; 0.85\}$ |
| Helical | CFD: $c_{n,t} = \{995.5, 775.2\}$ | $\gamma_{\{N,s;N,q;R,s\}}^{Helix,CFD} = \{2.24; 1.25; 1.09\}$ |
| TPW | From (7) and (8) | $\gamma_{\{N,s;N,q;R,r\}}^{TPW,SBT} = \{1.95; 6.75; -2.5\}$ |
| TPW | CFD: $c_{n,t} = \{995.5, 775.2\}$ | $\gamma_{\{N,s;N,q;R,r\}}^{TPW,CFD} = \{1.65; 8; -3\}$ |

*3.2.2 Validation of the hydrodynamic model*

The hydrodynamic model is validated with additional CFD simulations for different amplitudes and wavelengths than the ones used in the representative design for the esti-mation of interaction coefficients for the body and resistance coefficients for the tail. The study can be extended to other parameters such as the body radius, tail length, body type etc. Here, we considered only wavelength and amplitude for clarity and conciseness as design variables of flagellar propulsion. Moreover, frequency, the diameter of the body and fluid properties are lumped into the scaling Reynolds number used in simulations. Thus, for small Reynolds numbers, the velocity of the robot scales linearly with the frequency of tail rotations and its body size.

For swimmers with helical tails, hydrodynamic model results are compared with CFD simulation results in Figs. **4**a-f. Average forward velocity (Fig. **4**a), the amplitude of the lateral velocity (Fig. **4**b) and the body rotation rate (Fig. **4**c) are plotted against the amplitude, which is the radius of the helix. According to hydrodynamic model results with resistive force coefficients from Lighthill's slender body theory (SBT) [18], the magnitude of the time-averaged forward velocity increases with the amplitude with a rate that slows down at higher values. The model results with CFD-based force coefficients also show that the average velocity increases with the amplitude; in this case, a slightly better agreement with actual simulation results is observed that the case with SBT-based force coefficients. The agreement between the hydrodynamic surrogate model and simulation results is better at small values of the amplitude than large ones (Fig. **4**a), thus, indicating that as the helical radius increases and the flow induced by the tail gets stronger than the case used for the estimation of interaction coefficients the accuracy of the surrogate model deteriorates. Percentage errors from the plots are listed in Table 4 for all cases.

The time-dependent lateral motion of the swimmer is periodic with zero mean value. However, the amplitude of the lateral velocity increases with the amplitude of the helical waves almost linearly; the agreement is slightly better for the force coefficients from the slender body theory than the force coefficients obtained from the CFD simulation for the stationary swimmer (Fig. **4**b). Similarly, in Fig. **4**c, model results with analytically obtained force coefficients from the slender body theory agree with simulation results for large wave amplitudes better than the results with constant force coefficients (11.8% error vs. 42.7%); the agreement is poorer for both sets of coefficients at small amplitudes.

Average forward velocity, the amplitude of the lateral velocity and the average body rotation rate are plotted against the number of waves in Figs. **4**d-f, respectively. The forward velocity predicted by the hydrodynamic model indicates that the wavelength does not have a significant effect, and agrees well with CFD simulation results for both sets of parameters (Fig. **4**d) (6.9% for constant $c_n$ and $c_t$, and 9.3% for $c_n$ and $c_t$ from the slender body theory (SBT)). The amplitude of the lateral velocity peaks at half-integer values of the number of waves, i.e., for $N_\lambda = 1.5, 2.5, 3.5$, etc., and falls at full integer values. When the helical waves are in full-periods, forces in the lateral directions are minimal, and emerge only due to the bias introduced by the shape function $b_s(s)$ in Eq. (12). However, when the helical waves do not have full turns, the symmetry is broken and hydrodynamic forces in lateral directions emerge and the effect is maximized when the incomplete wave is half. Moreover, the intensity of the lateral motion diminishes as the number of waves increases indicating that the effect of the incomplete wave is diluted as the total number of waves increases. Overall, the hydrodynamic model predicts the lateral motion well especially with analytical resistive force coefficients compared to resistive force coefficients computed from the CFD simulation for the stationary swimmer (see Fig. **4**e). Lastly, the rotation rate of the body does not vary with the number of waves on the tail significantly and predicted reasonably well with the hydrodynamic model as shown in Fig. **4**f.

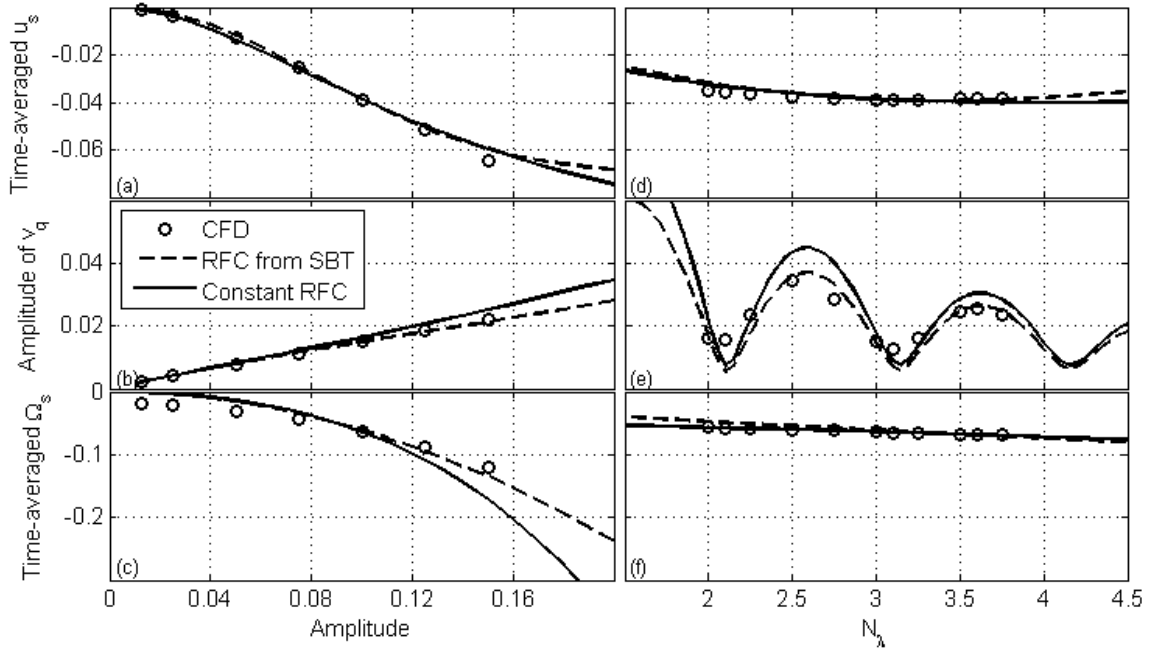

**Fig. 4.** Time-averaged forward (a,d), amplitude of the lateral (b,e), and rotation of the body (c,f) against the amplitude (a-c) and number of waves (d-f) for helical (RRH) tails: circles are CFD results, solid lines are for hydrodynamics model with resistive force coefficients obtained from the CFD simulation for a stationary swimmer, and dashed lines are for hydrodynamic model results with resistive force coefficients obtained from Lighthill's slender body theory (SBT) [18].

For swimmers with traveling-plane-wave tails, hydrodynamic model results are compared with CFD-simulation results in Figs. **5**a-f for both sets of parameters: resistive force coefficients from the slender body theory (RFC from SBT) and corresponding hydrodynamic interaction coefficients of the body; and resistive force coefficients determined from the CFD-simulation for the stationary swimmer with the helical tail (constant RFC) and corresponding interaction coefficients.

Time-averaged forward velocity, the amplitude of the lateral velocity and the amplitude of the angular velocity of the body in the **r-**direction increase with the amplitude, and are predicted very well with the hydrodynamic model for both sets of parameters (Figs. **5**a-c). The time-averaged forward velocity of the swimmer is plotted against the number of waves on the tail in Fig. **5**d. For small wave numbers, hydrodynamic model results agree well with CFD-simulation results for both sets of coefficients. However, time-averaged velocity calculated by the model with force coefficients from the slender body theory decreases with increasing number of waves for large values. Model results with constant force coefficients agree very well with CFD simulation results for large of waves numbers as well.

The lateral velocity of the swimmer in the **q-**direction is periodic in time with zero-average value. The amplitude of the lateral velocity varies with the number of waves on the tail similarly to helical tails with the exception that peaks are observed with the total number of waves being equal to full integers, and bottoms being equal to half-integers as shown in Fig. **5**e. In part this is because of the effect of the amplitude-shape function, which introduces a bias near the body and breaks the balance of forces towards the tip of the tail: for half-integer waves on the tail **q-**direction forces are symmetric and net force is small, on the other hand for full integer waves the motion of the tip of the tail is not balanced by the motion of the tail near the body. Hydrodynamic model results agree very well qualitatively with the CFD results despite a slight shift in the results for a total number of waves larger than 3.5. Moreover, the hydrodynamic model with resistive force coefficients from the slender body theory predicts that the overall trend of the amplitude of the lateral velocity decreases slowly with respect number of waves on the tail, although results of the model with constant force coefficients show a decreasing trend as the number of waves increases and agrees well with the CFD simulation results (see Fig. **5**e). Lastly, the amplitude of **r-**rotations of the body follows a trend with peaks near the half-integer waves and falls at slightly larger values than the full integer number of waves on the tail (see Fig. **5**f). Although the overall trend agrees well with the CFD simulation results, the range of the falls and peaks are not as large in CFD simulation results as in the hydrodynamic model results.

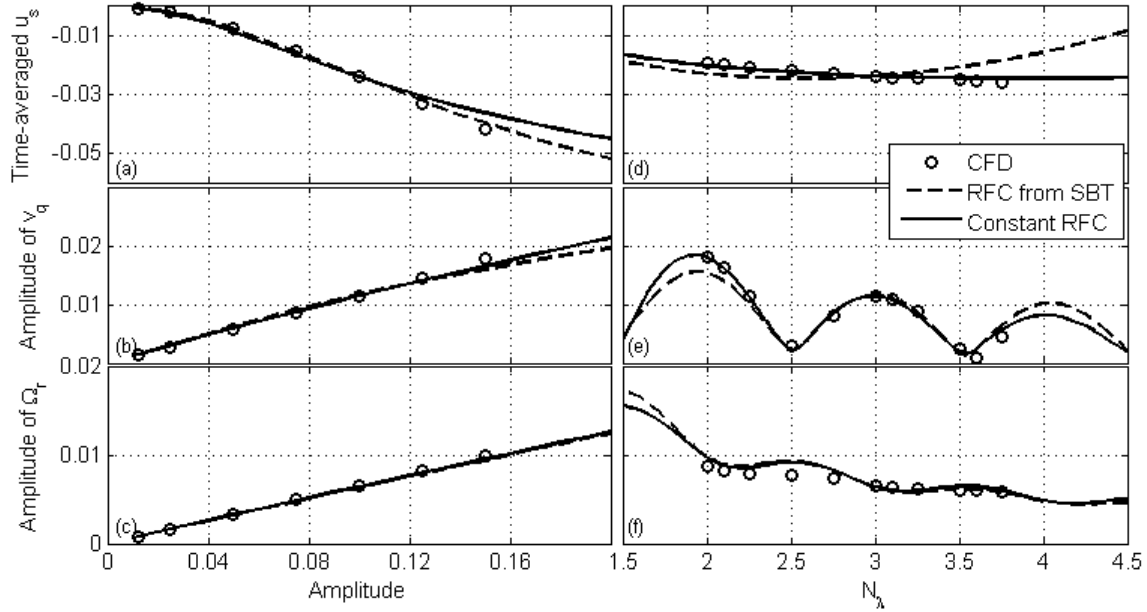

**Fig. 5.** Time-averaged forward velocity (a), amplitude of the lateral velocity (b), and amplitude of body rotation around the **r**-axis (c) are plotted against the amplitude of waves for TPW tails (a-c) and number of waves (d-f). Circles are CFD results, solid lines are for hydrodynamics model with resistive force coefficients obtained from the CFD simulation for a stationary swimmer with a helical tail, and dashed lines are for hydrodynamic model results with resistive force coefficients obtained from Lighthill's slender body theory (SBT) [18].

Summary of the performance of the hydrodynamic surrogate model is presented in Table 4. Overall, the surrogate model agrees very well with CFD simulation results for both sets of resistive force coefficients (RFC) used in the model and for both actuation types.

**Table 4**
Errors in predictions of the hydrodynamic model, (absolute error; range)

|  |  | Number of Waves | | Amplitude | |
|---|---|---|---|---|---|
|  |  | CFD-based constant $c_n, c_t$ | Analytical $c_n, c_t$ | CFD-based constant $c_n, c_t$ | Analytical $c_n, c_t$ |
| Helical tail | $\bar{u}_s$ | 0.0025, [-.0327,-.0393] | 0.0033, [-.0319,-.03939] | 0.0047, [-.0011,-.0596] | 0.0046, [-.0008,-.0597] |
|  | $v_{q,\max}$ | 0.0085, [.0073,.0435] | 0.0102, [.0052,.0356] | 0.003, [.002,.0251] | 0.00074, [.002,.0212] |
|  | $\bar{\Omega}_s$ | 0.0011, [-.0574,-.0694] | 0.00843, [-.0708,-.4783] | 0.0516, [-.0009,-.1722] | 0.00143, [-.0009,-.1349] |
| Traveling plane waves | $\bar{u}_s$ | 0.0011, [-.020,-.024] | 0.0031; [-.0186, -.0226] | .0056; [-.0006,-.0364] | 0.0023; [-.0005,-.0397] |
|  | $v_{q,\max}$ | .0013; [.0023, 0.018] | 0.0022; [.0015, .0203] | 0.0012; [.0015, .0165] | 0.0017; [.0016, .0160] |
|  | $\Omega_{r,\max}$ | 0.0014; [.0059,.0096] | 0.0010; [.0058, .0095] | $4.1 \times 10^{-4}$; $[8.2, 95.1] \times 10^{-4}$ | $5.5 \times 10^{-4}$; $[8.1, 93.8] \times 10^{-4}$ |

## 4. Applications of hydrodynamic model

### 4.1 Design with the hydrodynamic model

Design of an artificial micro swimmer can be carried out with the validated hydrodynamic surrogate model that can replace the computationally exhaustive three-dimensional CFD model. For example, energy consumption of the robot, for which the base case parameters are given in Table 1, can be minimized with the maximization of its efficiency, which is given by:

$$\eta = \frac{\Pi_{body}}{\Pi_{tail}} \tag{24}$$

where $\Pi_{body} = F_s u_s$ is the average rate of work done to move the body of the robot with the velocity of $u_s$ against the drag force on the body, $F_s$, and $\Pi_{tail}$ is the rate of work done to actuate the tail of the robot and calculated from $\Pi_{tail} = T_s \omega$ for helical tails, where $T_s$ is the torque needed to rotate the tail with angular velocity, $\omega$. For traveling plane waves, the rate of actuation work is calculated from the integration of the product of the local force and the local net velocity in the lateral direction, i.e., $F_q(s) \cdot dq(s,t)/dt$, over the entire tail length.

Average forward velocity (Fig **6**a,b) and the hydrodynamic efficiency of swimmers (Fig. **6**c,d) are calculated with the hydrodynamic model for amplitudes varying between .01 and .5 and for a total number of waves between 0.5 and 5. According to Figs. **6**a and **6**b, there is a similarity between the forward velocity of swimmers with helical tails and traveling plane waves, former with the maximum velocity of 0.21 for $b = 0.5$ and $N_\lambda = 1$, and the latter with the maximum velocity of 0.12 for the same amplitude and $N_\lambda = 0.8$. Therefore, in order to design a swimmer with the fastest velocity, one has to build a tail with a single helical turn with the largest amplitude. Moreover, swimmers with helical tails are considerably faster than the ones with traveling plane waves.

From Figs. **6**c and **6**d, the efficiency of the swimmers with helical tails are considerably larger than the efficiency of the swimmers with traveling plane waves; the maximum efficiency for the helical tails is obtained as 2.5%, and as 0.29% for traveling plane waves for the robots with geometric parameters as given in Table 3.

In addition to geometric design, the hydrodynamic model can also be used to estimate physical properties of natural swimmers. Similar to the procedure of obtaining the interaction coefficients through the solution of the inverse problem discussed above, geometric properties and wave propagation parameters of a natural swimmer can be determined from the hydrodynamic model. For example, given the swimming trajectory of a particular spermatozoa specimen, e.g., the bull sperm cells studied by Friedrich et al. [48], the cor-responding wave shape and pattern can be obtained from the solution of the inverse problem. Moreover, Gurarie et al. [49] demonstrated that stochastic model can be used for the prediction of the full three-dimensional trajectory of the swimmer based on two-dimensional observations; hydrodynamic models can be used to improve the predictability of complex trajectories.

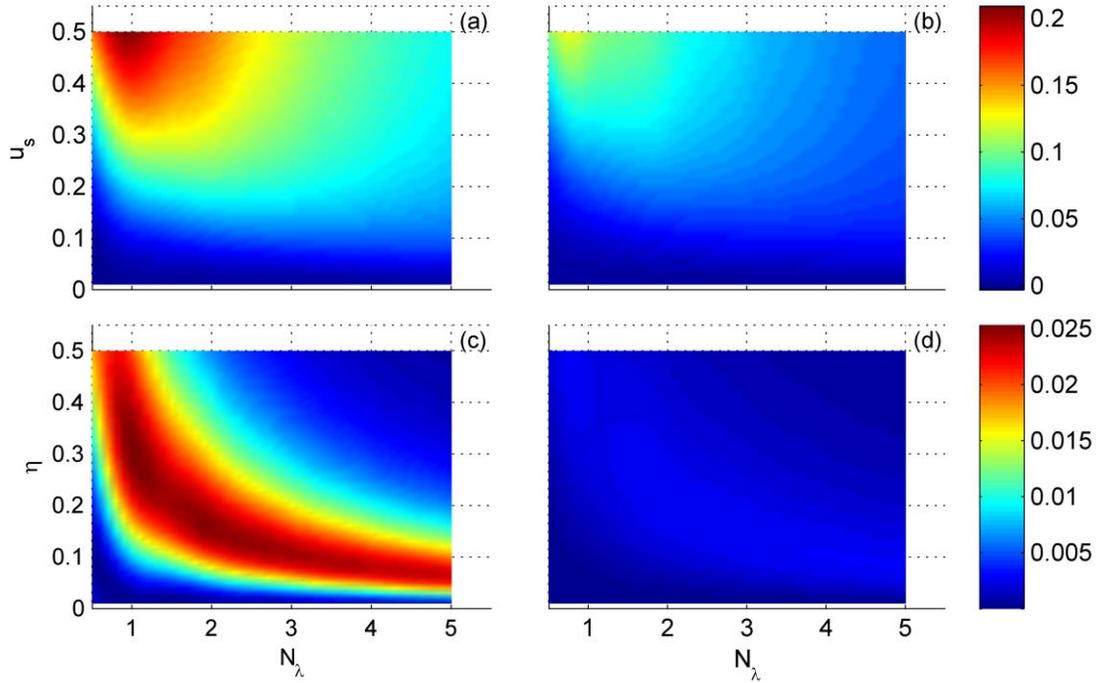

**Fig. 6.** Time-averaged forward velocity, $u_s$ (a,b) and hydrodynamic efficiency, $\eta$ (c,d); obtained by the hydrodynamic model for robots with helical tails and traveling plane waves, respectively.

## 5. Conclusions

Forward and lateral translational and rotation of bio-inspired micro swimmers that consist of a body and an actuated tail are predicted with a hydrodynamic surrogate model, which is based on a number of parameters used in the resistive relationship between the force and velocity vectors on the tail and the body. The hydrodynamic model runs essen-tially in real-time to predict the full trajectory of swimmers, unlike the three-dimensional CFD model that completes the solution of the Navier-Stokes equations in hours if not a day.

For the actuated tail of the swimmer, which is considered as either a rotating helix or traveling-plane-wave deformations on a flexible rod, two sets of resistive force coefficients are used: one set is from the slender body theory of Lighthill [18], and the second set is directly calculated from a single CFD simulation for a stationary swimmer with a helical tail for which the amplitude and wavelength are set to 0.1 and 2/3, respectively in non-dimensional units. For each form of flagellar actuation and the set of force coefficients, hydrodynamic interaction coefficients are estimated for the body of the swimmer from the solution of the inverse problem for the base case values of the amplitude and the wavelength. Then the hydrodynamic model is validated directly against CFD model results for swimmers with helical and traveling-plane-wave tails for which the amplitude is varied between 0.01 and 0.15 and the wavelength is varied between 0.5 and 1. For all cases, the surrogate hydrodynamic model results agree reasonably well with CFD model results.

Furthermore, experimentally measured time-averaged forward velocity and body rotation rates for microorganisms that are presented in the literature are compared with the results of the hydrodynamic model with resistive force coefficients obtained from the slender body theory. Once the hydrodynamic interaction coefficients of the body are determined from the inverse problem for a fixed specimen, predicted forward velocities and body rotation rates agree very well with the measurements for other species with the different body and tail dimensions.

Lastly, we demonstrated the application of validated hydrodynamic surrogate models in the design of bio-mimetic robots to obtain optimal propulsion type, amplitude, and wavelengths. Moreover, surrogate hydrodynamic models can be used to determine geometric properties of natural swimmers from their observed trajectories with the rapid turn-around in solutions of the inverse problem.

**Appendix**

$A_1$ and $A_2$ periodic integrals as articulated by Lighthill [18]:

$$\phi = \left[\alpha^2 \theta^2 + 2(1-\alpha^2)(1-\cos(\theta))\right]^{3/2}, \tag{25}$$

$$\varepsilon = 5.2 r_t \alpha / \lambda, \tag{26}$$

$$A_1(\alpha) = \ln \varepsilon + \int_\varepsilon^\infty \theta \sin(\theta)/\phi \, d\theta, \tag{27}$$

$$A_2(\alpha) = \ln \varepsilon + \int_\varepsilon^\infty \sin^2(\theta)/\phi \, d\theta. \tag{28}$$

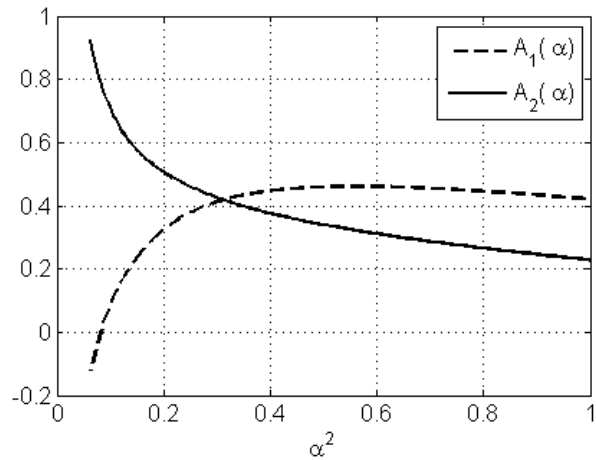

**Fig. 7.** $A_1(\alpha)$ and $A_2(\alpha)$ with respect to $\alpha^2$.

**Acknowledgements**

Financial support for this study was provided in part by a grant from The Scientific and Technological Research Council of Turkey.